\documentclass[pra,twocolumn]{revtex4}
\usepackage[english]{babel}
\usepackage{amsmath}
\usepackage{amssymb}
\usepackage{epsfig}
\usepackage{url}

\begin{document}
\title{Small-Scale Light Structures in a Kerr Medium}
\author{Victor P. Ruban}
\email{ruban@itp.ac.ru}
\affiliation{Landau Institute for Theoretical Physics RAS,
Chernogolovka, Moscow region, 142432 Russia}

\date{\today}

\begin{abstract}
A system of equations has been proposed for a monochromatic weakly 
nonlinear light wave in a Kerr medium. This system is equivalent up
to the third order in electric field to the known equation
$\mbox{curl}\,\mbox{curl}\, {\bf E}=k_0^2[{\bf E}
+\alpha |{\bf E}|^2{\bf E}+\beta({\bf E}\cdot{\bf E}){\bf E}^*]$, 
but the new equations are much more convenient for numerical computation.
Optical fields with small structures of two or three wavelengths have been 
simulated using this system. It has been found that a stable self-focused 
light beam (a 2D vector soliton) in some parametric domain is possible 
even without modification of nonlinearity. ``Inelastic'' collisions between 
two such narrow beams with opposite circular polarizations have been computed. 
Furthermore, examples of interacting optical vortices, spatial separation 
of the circular polarizations, and the Kelvin--Helmholtz instability have 
been given for defocusing nonlinearity.
\end{abstract}

\maketitle

\section{INTRODUCTION}
As known, a monochromatic light wave in a transparent nonlinear optical
Kerr medium is described by the vector nonlinear equation [1]
\begin{equation}
\mbox{curl}\, \mbox{curl}\, {\bf E}=k_0^2[{\bf E}
+\alpha |{\bf E}|^2{\bf E}+\beta({\bf E}\cdot{\bf E}){\bf E}^*],
\label{curl_curl}
\end{equation}
where ${\bf E}(x,y,z)$ is the complex amplitude of the fundamental 
harmonic of the electric field, $k_0$ is the wavenumber, and $\alpha$ 
and $\beta$ are the nonlinear coefficients ($\beta=\alpha/2$ in the limit
of the instantaneous nonlinear response). This equation in principle
allows one to examine numerous phenomena in a propagating wave at small
scales, e.g., the internal spatial dynamics of intense narrow light beams
and their interaction with each other in both polarizations. Another
obvious application is to study nonlinear effects near the focus of a 
nonparaxial (i.e., collecting light at a quite wide angle) lens. 
Unfortunately, Eq.(1) is insufficiently convenient if one of the Cartesian
coordinates is considered as an evolution variable. In particular, 
the system for waves propagating on average along the $z$ axis does not
contain the evolution derivative $\partial_z E_z$ or $\partial_z^2 E_z$, 
but the longitudinal field component satisfies a complicated elliptic 
equation in the transverse coordinates. This difficulty stimulates the
search for alternative approaches (see, e.g., [2--9] and references therein).
The most popular approximation is the system of coupled nonlinear Schr\"odinger
equations, where the amplitudes $A_{1,2}(x,y,z)$  of the left and right
circular polarizations in the expression
\begin{equation}
{\bf E}\approx \big[({\bf e}_x+i{\bf e}_y) A_1 
+ ({\bf e}_x-i{\bf e}_y) A_2 \big]\exp(ik_0 z)/\sqrt{2} 
\end{equation}
are assumed to be slowly varying functions. When the longitudinal component
is completely neglected, a pair of nonlinear Schr\"odinger equations are
obtained in the form [2]
\begin{equation}
-2i k_0 \partial_z A_{1,2}=\Delta_\perp A_{1,2} 
+\alpha k_0^2\Big[|A_{1,2}|^2 + g|A_{2,1}|^2\Big]A_{1,2},
\label{A_12_eqs}
\end{equation}
where $g=1+2\beta/\alpha$ is the cross phase modulation parameter. 
This system corresponds to two interacting quantum fluids with the densities
$I_{1,2}=|A_{1,2}|^2$, and the amount of each of them is conserved along the 
$z$ axis. An analogue between light and a bimodal quantum fluid is fairly deep
(see, e.g., [3--5] and references therein).

In the case of focusing nonlinearity (positive coefficients $\alpha$ and 
$\beta$), Eqs.(3) provide the well-known conclusion on the impossibility
of stable two-dimensional solitons: a wave packet either spreads due to
diffraction or collapses due to nonlinearity (see [10, 11] and references 
therein). The saturation of nonlinearity or even change in its sign at high
field intensities (see, e.g., [12--15]) are usually considered as mechanisms
preventing the collapse at small scales about the wavelength $\lambda_0=2\pi/k_0$.
However, that consideration is beyond the third order in the amplitude. 
In the third order, nonparaxial corrections due to terms neglected in Eqs.(3)
can be taken into account. The nonlinear Schr\"odinger equations thus modified
are quite complicated and insufficiently reliable and can be solved only 
numerically [6--8]. Consequently, to study small-scale structures, it is
reasonable to focus on models nonperturbative in transverse wavenumbers.

In this work, a simple method is proposed to overcome difficulties caused
by the vector nature of Eq.(1). A system of equations convenient for 
numerical simulation, which differs from the initial Eq.(1) only beginning
with the fifth order in the field amplitude, is presented. Computer experiments
with this model demonstrate the existence of periodically self-focusing optical
beams (see examples in Figs.1,2). It is noteworthy that such a behavior has 
been already observed in some numerical experiments with less accurate 
models [7--9]. In the limit of low amplitude of radial oscillations, these 
beams are stable two-dimensional solitons (thin light beams uniform in the $z$ 
coordinate), which are balanced due to nonlinear diffraction. A number of other
interesting numerical examples demonstrating the capabilities of the new 
approach are also given.

\section{MODEL}

First, it is known that change of the dependent variables 
${\bf E}={\bf u}+\nabla(\nabla\cdot{\bf u})/k_0^2$ in the linear case splits
the vector equation  $\mbox{curl}\,\mbox{curl}\, {\bf E}=k_0^2{\bf E}$
into three scalar Helmholtz equations for the components of the vector ${\bf u}$
because the term $\nabla(\nabla \cdot {\bf u})$ cancels. The idea is to use
a similar substitution 
\begin{equation}
{\bf E}=U_1{\bf e}_x+U_2{\bf e}_y +\nabla(\partial_x U_1+\partial_y U_2)/k_0^2 
+\nabla\phi 
\label{U_phi}
\end{equation}
for the nonlinear Eq.(1), where the components $U_1(x,y,z)$ and $U_2(x,y,z)$ 
of the transverse vector ${\bf U}$ are of the first order, whereas the additional
nonlinearity-induced scalar potential $\phi(x,y,z)$ is a third-order term. 
For the sake of brevity, the following notation is introduced
\begin{equation}
{\bf F}={\bf U}+\nabla_\perp(\nabla_\perp \cdot {\bf U})/k_0^2
+{\bf e}_z(\nabla_\perp \cdot \partial_z{\bf U})/k_0^2.
\label{F_def}
\end{equation}
The field ${\bf F}$ in the leading order coincides with ${\bf E}$, so that the
three-dimensional vector structure of the electric field is primarily determined
by only two scalar functions $U_1(x,y,z)$ and $U_2(x,y,z)$ (including the
longitudinal component $E_z$ given by the last term in Eq.(5)). 

It is important that the function $\phi$ in nonlinear terms can be omitted when
substituting Eq.(4) into Eq.(1) because the corresponding terms are of the fifth
order and higher. As a result, the transverse vector ${\bf U}$ satisfies the
equation
\begin{equation}
-(\partial_z^2+\Delta_\perp){\bf U}=k_0^2[{\bf U}+\nabla_\perp\phi
+\alpha|{\bf F}|^2{\bf F}_\perp 
+\beta ({\bf F}\cdot{\bf F}){\bf F}^*_\perp].
\label{U_eq}
\end{equation}
The first-order terms do not remain in the longitudinal component of Eq.(1): 
\begin{equation}
-\partial_z \phi=\alpha|{\bf F}|^2 F_z
+\beta ({\bf F}\cdot{\bf F}) F^*_z.
\label{phi_eq}
\end{equation}
Equations (5)--(7) specify our model, where nonlinear diffraction effects 
are due to derivatives in Eq.(5).

\begin{figure}
\begin{center} 
\epsfig{file=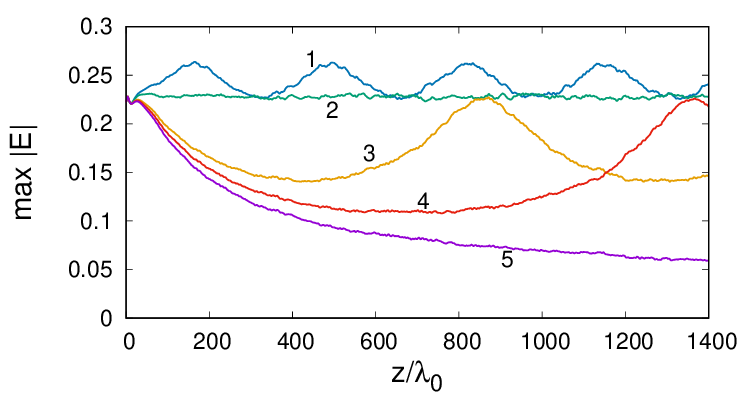, width=86mm}
\end{center}
\caption{
Maximum field amplitude under repeated self-focusing of the light beam 
with several initial widths (see details in the main text) versus 
the $z$ coordinate.
}
\label{beams} 
\end{figure}

\begin{figure}
\begin{center} 
\epsfig{file=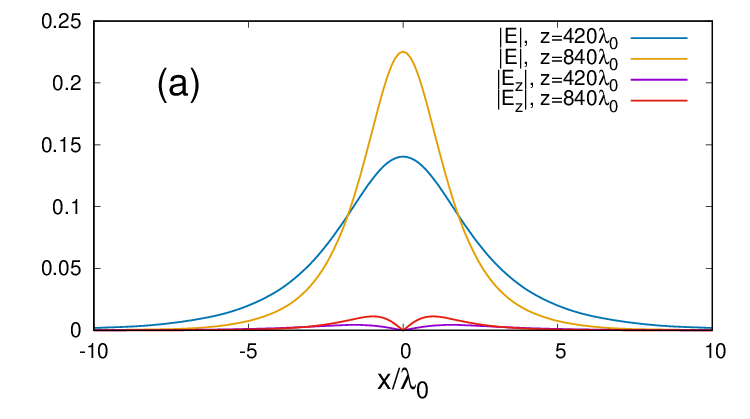, width=86mm}\\
\epsfig{file=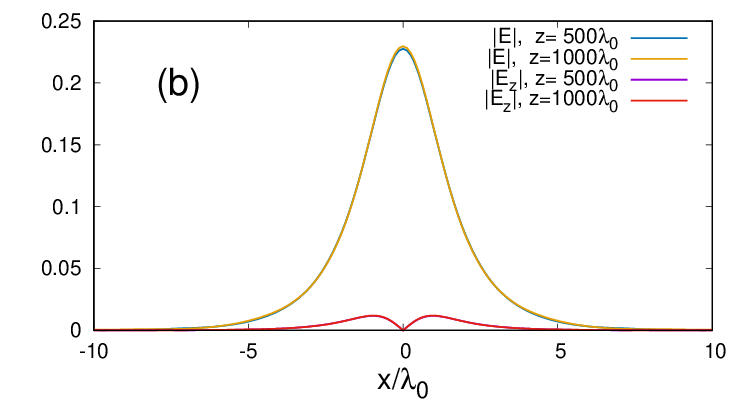, width=86mm}
\end{center}
\caption{
Field amplitude profiles (a) at the minimum and maximum focusing of the 
periodically self-focusing beam corresponding to line 3 in Fig.1 and 
(b) in two sections of the approximately balanced beam corresponding to 
line 2 in Fig.1 (in this case, plots for different $z$ values almost 
coincide with each other). 
}
\label{refocusing} 
\end{figure}

\section{NUMERICAL METHOD}

Since each of Eqs.(6) and (7) is resolved with respect to the corresponding
highest derivative over the variable $z$, this system allows a quite simple 
numerical simulation by the pseudospectral method. However, nonlinear equations
at transverse wave vectors ${\bf k}_\perp$ exceeding $k_0$ in absolute value
should be regularized because the dispersion law $k_z^2=(k_0^2-|{\bf k}_\perp|^2)$
at high $|{\bf k}_\perp|$ values does not correspond to propagating waves. 
In this case, the dispersion law was regularized as
\begin{equation}
(k_0^2-|{\bf k}_\perp|^2) \to k_0^2-|{\bf k}_\perp|^2
[1+|{\bf k}_\perp|^8/k_0^8]^{-1/4}. 
\end{equation}
Furthermore, after each step of the integration of the system in the $z$
coordinate by the fourth-order Runge–Kutta method, all spectral components 
near $k_0$ and further were filtered with a specially selected function 
$f(|{\bf k}_\perp|)$, decreasing from unity at $|{\bf k}_\perp|\approx 0.9 k_0$ 
to zero at some $K\approx 1.5 k_0$. After this procedure, transverse spectra
remain quite rapidly (about exponentially) decreasing; at $|{\bf k}_\perp|$ 
near $k_0$, they are three to four orders of magnitude smaller than the main
spectral components. It is also natural that nonlinear spectral ``expansion''
is slower at a weaker nonlinearity (at the field scale corresponding to the 
coefficient $|\alpha|=1$, nonlinearity becomes strong at $|E|\sim 1$; 
the maximum field in our calculations reached 0.5--0.6, which still corresponds
to moderate nonlinearity; it is a different problem whether this intensity is
acceptable for a real physical medium and whether it initiates processes 
disregarded in the model specified by Eq.(1) [16]).

The computational domain was a $2\pi\times 2\pi$ square with periodic boundary
conditions and dimensionless $k_0=40$ was taken. The discretization step in the
transverse coordinates was $h_\perp=2\pi/320$, i.e., eight numerical lattice
points per the wavelength $\lambda_0$. The step in the longitudinal coordinate
in different series of experiments was $h_z=0.01\lambda_0$ or $h_z=0.0004$. 
The nonlinear coefficients for the focusing (defocusing) medium are set
$\alpha=1$ and $\beta=0.5$ ($\alpha=-1$ and $\beta=-0.5$).

\section{TWO-DIMENSIONAL SOLITONS}

Of great interest is the possible balance between diffraction and nonlinearity
at small scales, where nonparaxial effects become noticeable and the nonlinear
Schr\"odinger equations are no longer applicable. To simulate the behavior of
such narrow light beams, we took the initial configuration in the form of the
circularly polarized wave
\begin{equation}
{\bf U}(x,y,0)=({\bf e}_x+i{\bf e}_y) S(x,y,w), 
\end{equation}
where the profile $S(x,y,w)$ is approximately the sum of two Gaussians with
sufficiently large parameters $w\sim 10$: 
\begin{eqnarray}
S(x,y,w)&=&0.10\exp(-40[R(x)+R(y)])\nonumber\\
        &+&0.06\exp( -w[R(x)+R(y)]).
\end{eqnarray}
where $R(\xi)=(1-\cos(\xi))[1+(1-\cos(\xi))/6]\approx  \xi^2/2$. Since the
parameters in this expression were selected roughly after only a few trial
simulations, it should not be treated too seriously. 

It is noteworthy that the solutions of our model, as should be, can include both
the direct, [$\sim\exp(ik_0 z)$], and opposite, [$\sim\exp(-ik_0 z)$], waves. 
To make the opposite wave negligibly weak, the initial derivative was 
$\partial_z{\bf U}(x,y,0)=1.01 i k_0 {\bf U}(x,y,0)$. The initial potential 
$\phi(x,y,0)$ was set to zero.

Figure 1 shows the dependence of the maximum electric field amplitude on the $z$
coordinate at $w=7.9, 8.0, 8.2, 8.25$, and $8.3$ (the corresponding lines from 
1 to 5). It is seen that the curves are similar to strongly anharmonic 
one-dimensional oscillations of a certain fictitious particle in an asymmetric
potential, except for the last aperiodic case. This behavior corresponds to the
multiple repeated self-focusing of the light beam. An example of field amplitude
profiles in the minimum and maximum is presented in Fig.2a. We also note that
radial oscillations at $w=8.0$ are so small that the beam can be treated as 
a balanced $z$-uniform two-dimensional soliton, which is confirmed in Fig.2b. 

It is quite probable that there is a Hamiltonian system close to our model,
where such vector solitons are stable extrema. The search for such a conservative
system is problem for future.

\begin{figure}
\begin{center} 
\epsfig{file=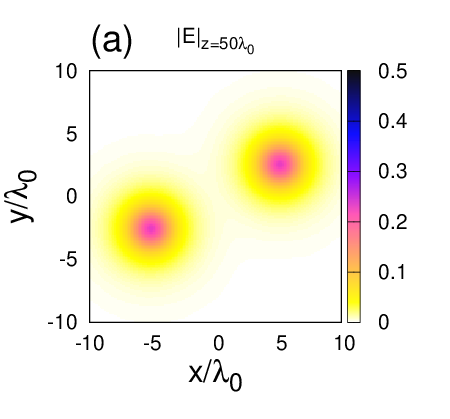, width=42mm}
\epsfig{file=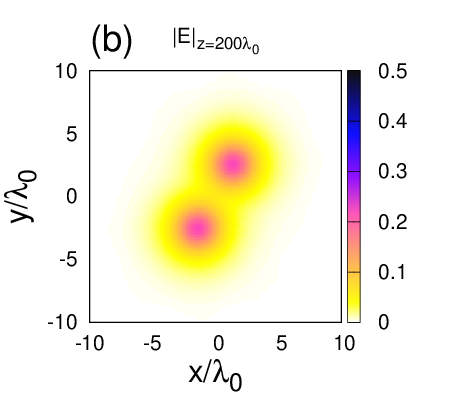, width=42mm}\\
\epsfig{file=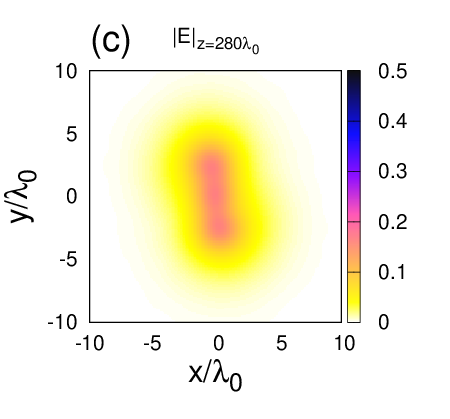, width=42mm}
\epsfig{file=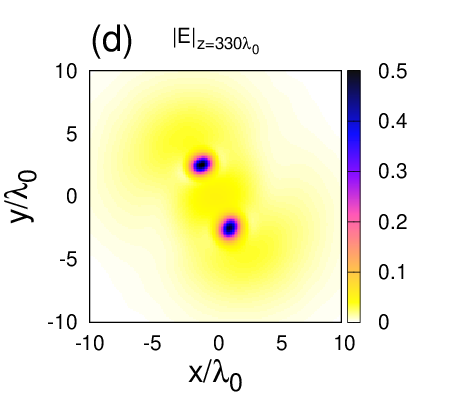, width=42mm}\\
\epsfig{file=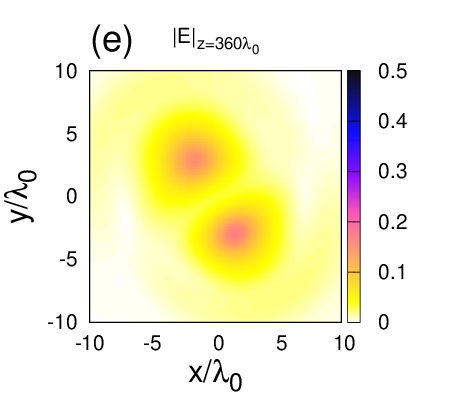, width=42mm}
\epsfig{file=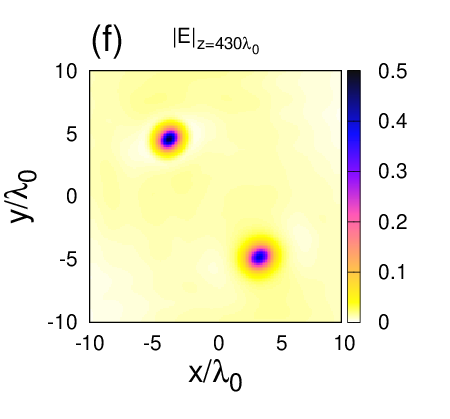, width=42mm}
\end{center}
\caption{
Example of the tangential collision of two thin light beams with 
opposite circular polarizations. Only the central part of the 
computational domain is shown.
}
\label{tangent} 
\end{figure}

It is of interest to simulate the interaction of thus obtained stable (at least, 
long-lived) light beams at different initial positions and directions with respect
to the $z$ axis. The tangential collision of two beams with opposite circular
polarizations is exemplified in Fig.3 (where an ``invert'' color palette is used 
to clearly visualize regions with low field amplitudes). In this case, 
the initial profile was 
\begin{eqnarray}
{\bf U}(x,y,0)&=&({\bf e}_x+i{\bf e}_y) S(x-1.0,y-0.4,8.0)e^{-ix}\nonumber\\
              &-&(i{\bf e}_x+{\bf e}_y) S(x+1.0,y+0.4,8.0)e^{ix}.
\end{eqnarray}
Two interacting solitons partially exchanged their polarizations. As a result, 
two slightly different beams were scattered from the place of collision with
approximately linear polarizations in a strongly oscillatory regime at a certain
scattering angle depending on the collision velocity and the impact parameter. 
The mentioned slight difference between scattered beams is due to the absence of
symmetry in the initial condition and indicates effects beyond the nonlinear
Schr\"odinger equations. A small energy fraction at collision was emitted in 
a weak ``free'' wave. Figure 4 shows the $z$-coordinate dependences of the maximum
field amplitude (a) in this numerical experiment and (b) in the case of initially
parallel close beams, where the initial state was specified by the function
\begin{eqnarray}
{\bf U}(x,y,0)&=&({\bf e}_x+i{\bf e}_y) S(x-0.4,y-0.4,8.0)\nonumber\\
              &-&(i{\bf e}_x+{\bf e}_y) S(x+0.4,y+0.4,8.0).
\end{eqnarray}
It is worth noting that the inelasticity of collisions is particularly pronounced
just in the case of parallel initial beams, because they are scattered after
collision at finite angles. 

A qualitatively similar partial exchange of polarizations between colliding 
three-dimensional light bullets was recently observed in [17].

\begin{figure}
\begin{center} 
\epsfig{file=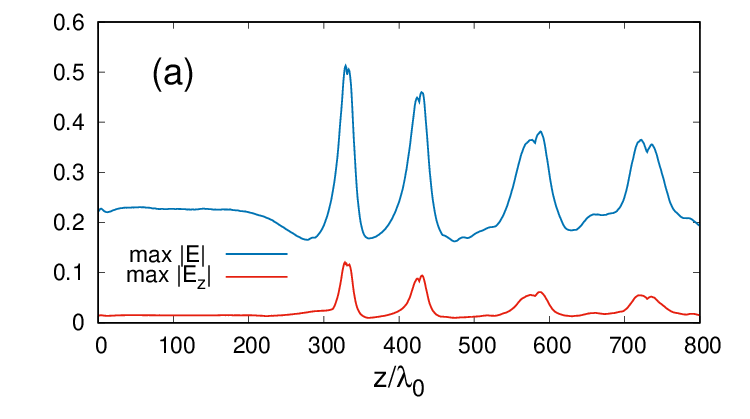, width=86mm}
\epsfig{file=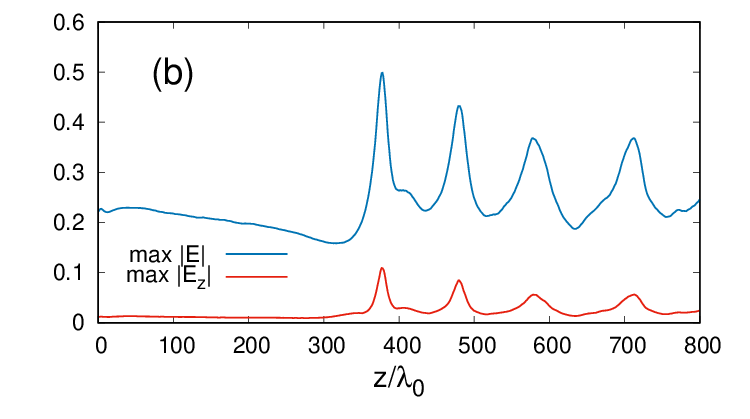, width=86mm}
\end{center}
\caption{
Maximum field amplitude versus the $z$ coordinate for two beams with opposite
circular polarizations for (a) the tangential collision shown in Fig.3 and
(b) initially parallel beams. 
}
\label{collisions} 
\end{figure}

The initial intensity in all above examples was already high enough and the 
transverse scale was sufficiently small so that a multiple amplification of 
the wave did not occur. Our method is also obviously applicable to simulate 
the strong relative focusing when the wave intensity increases by several orders
of magnitude. This requires only a higher spatial resolution and higher $k_0$ 
values. Strong nonlinear focusing of light behind a ``bad'' lens is exemplified
in Fig.5 for $k_0=60$. In this numerical experiment, the electric field magnitude
increases by more than an order of magnitude and, correspondingly, the intensity,
by two orders of magnitude. Systems with stronger relative focusing can be 
simulated in two stages. The initial light propagation occurring in the linear
regime can be evaluated with the theory of linear diffraction. Then, before the
beginning of the nonlinear regime, the obtained linear solution should be used
as the initial state for strongly nonlinear computations.

\begin{figure}
\begin{center} 
\epsfig{file=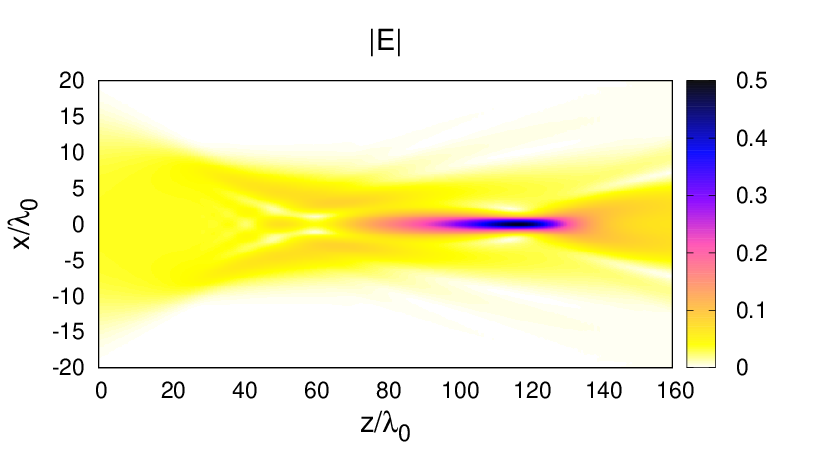, width=86mm}
\end{center}
\caption{
Field amplitude distribution in the longitudinal $y=0$ plane in the case of
nonlinear focusing of the circularly polarized wave behind the imperfect lens
whose central part almost does not deflect rays. The focused structure has
the shape of a thin needle.
}
\label{lens} 
\end{figure}

\section{STRUCTURES IN THE DEFOCUSING MEDIUM}

\begin{figure}
\begin{center} 
\epsfig{file=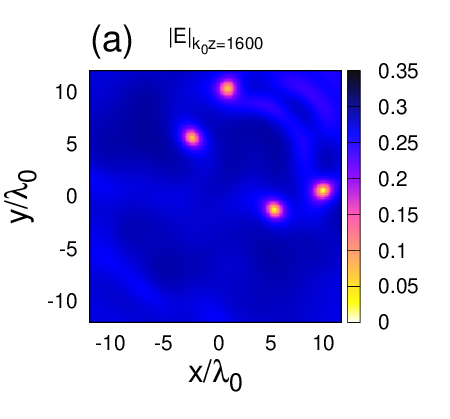, width=42mm}
\epsfig{file=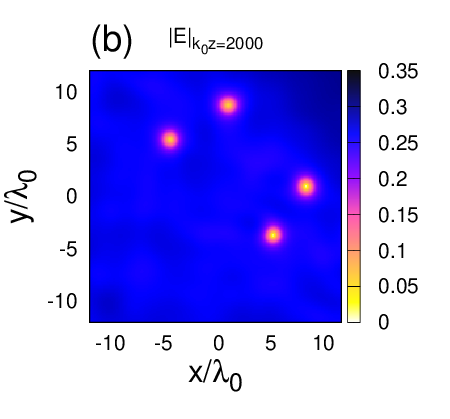, width=42mm}\\
\epsfig{file=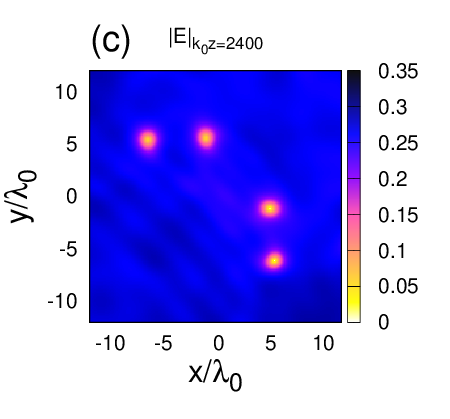, width=42mm}
\epsfig{file=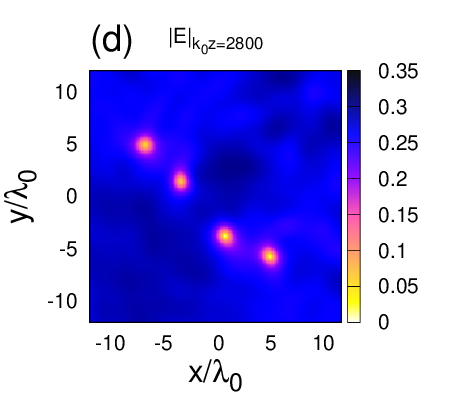, width=42mm}
\end{center}
\caption{
Example of interacting optical vortices in the circularly polarized wave in 
the presence of defocusing nonlinearity.
}
\label{vortices} 
\end{figure}

\begin{figure}
\begin{center} 
\epsfig{file=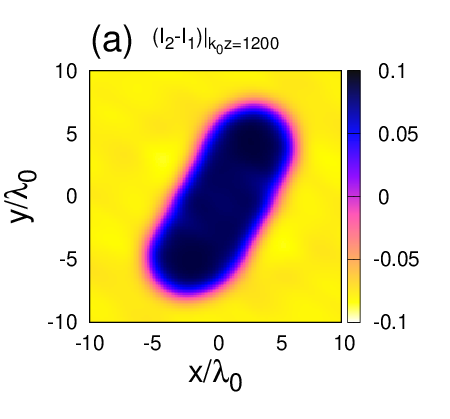, width=42mm}
\epsfig{file=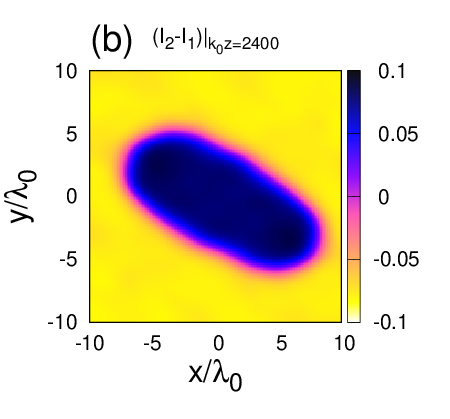, width=42mm}\\
\epsfig{file=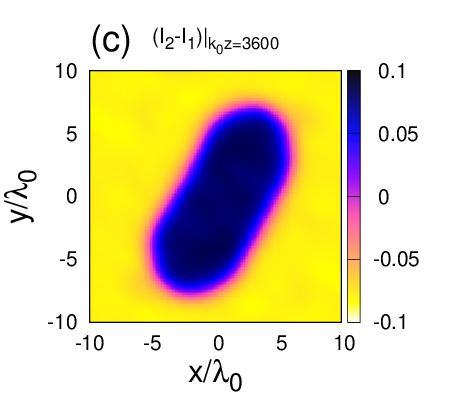, width=42mm}
\epsfig{file=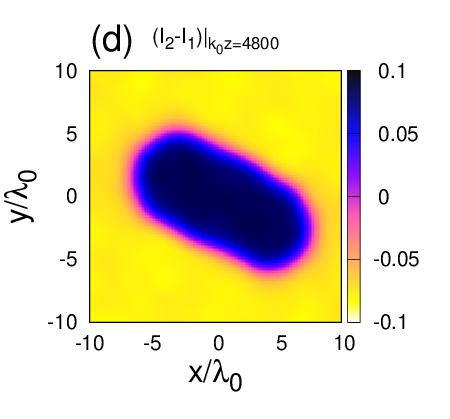, width=42mm}
\end{center}
\caption{
Example of an optical bubble in the defocusing medium. Quadrupole oscillations
of the shape of the domain wall occur with increasing $z$.
}
\label{bubble} 
\end{figure}

\begin{figure}
\begin{center} 
\epsfig{file=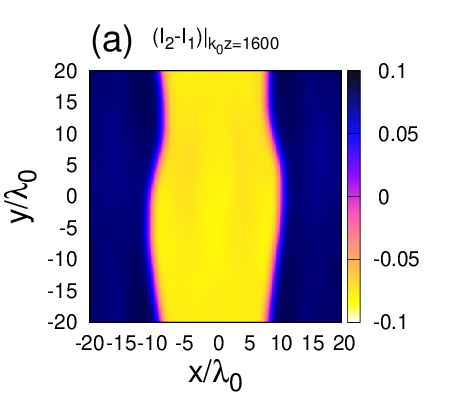, width=42mm}
\epsfig{file=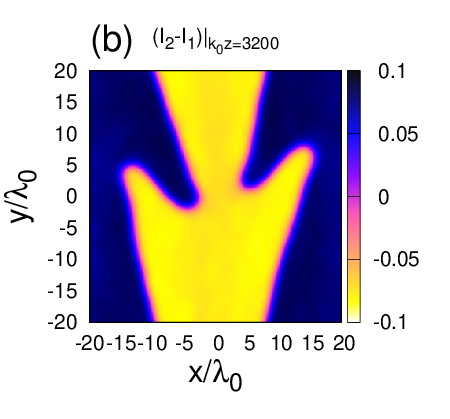, width=42mm}\\
\epsfig{file=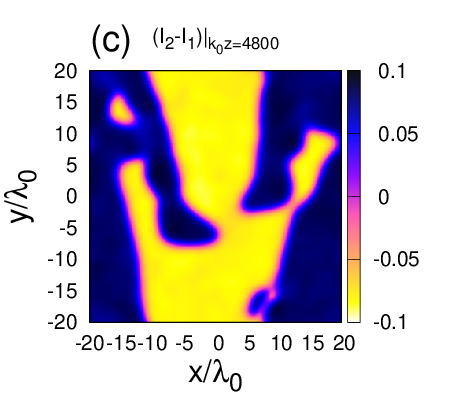, width=42mm}
\epsfig{file=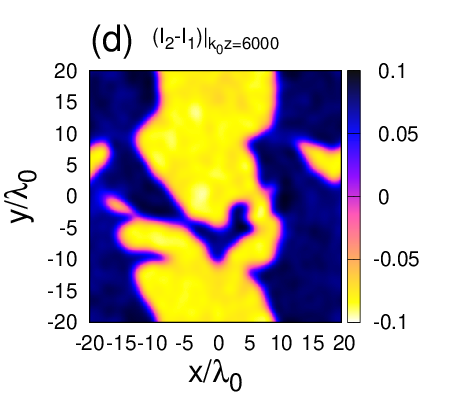, width=42mm}
\end{center}
\caption{Example of the development of the Kelvin--Helmholtz instability.
}
\label{KHI} 
\end{figure}

Several numerical examples, without details, are also given below for the 
defocusing case. Small-scale structures against a nearly uniform intensity
background are of interest in the case of such nonlinearity. Optical vortices
are the most known of them (see, e.g., [18--24] and references therein). 
Since circular polarizations are basic in the long-scale approximation (coupled 
nonlinear Schr\"odinger equations), just a circularly polarized wave was used as
the background in our numerical experiments with vortices. An example of two 
vortex--antivortex pairs is presented in Fig.6. It is seen that the cores of
vortices are as small as about two wavelengths against a background amplitude
of about 0.3.

A domain wall between two opposite circular polarizations is another interesting
object in the defocusing medium [25--31]. According to the coupled nonlinear
Schrödinger equations, this structure can exist under the condition 
$\beta/\alpha > 0$, which is satisfied in our case. Figure 7 exemplifies 
an optical bubble whose boundary is domain wall. Since the initial state was
not equilibrium, the bubble oscillates due to effective ``surface tension'' [32].

If the spatially separated circular polarizations move tangentially along the domain
wall relative to each other, the Kelvin--Helmholtz instability in its ``quantum'' 
variant is possible [33--36]. Successive stages of such an instability, with the
initial state in the form of slightly distorted ``sliding'' bands of the left and
right polarizations, are seen in Fig.8.

The development of the instability of the linearly polarized wave is shown in 
the last example in Fig.9. The initial state included spatially unseparated
left- and right-polarized waves each with a pair of ``bare'' vortices. As $z$
increases, the opposite polarization begins to fill the core of each vortex, 
specific modulation waves appear, and a quasirandom pattern of domains suggestive
of a geographical map is then formed.

\begin{figure}
\begin{center} 
\epsfig{file=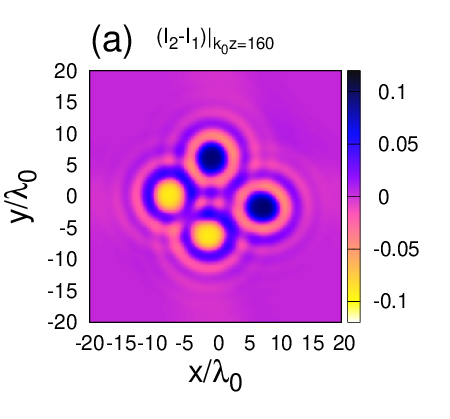, width=42mm}
\epsfig{file=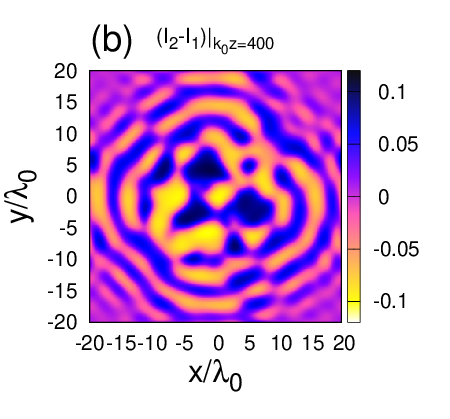, width=42mm}\\
\epsfig{file=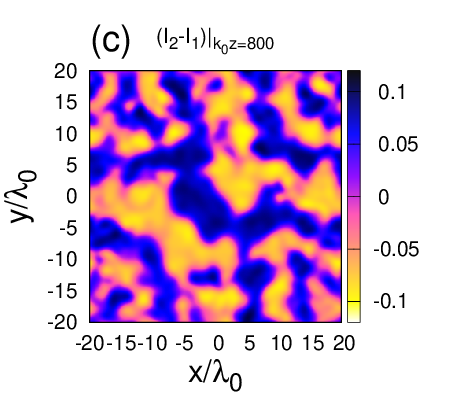, width=42mm}
\epsfig{file=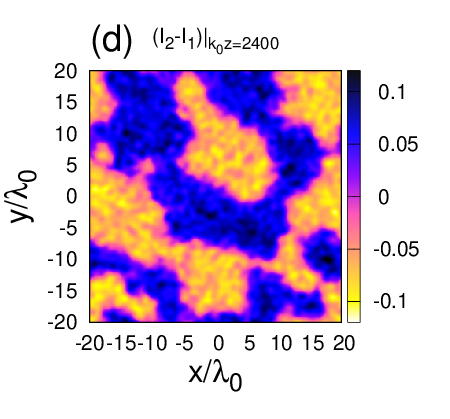, width=42mm}
\end{center}
\caption{
Example of the development of the instability of a linearly polarized wave 
in the defocusing medium. As the $z$ coordinate increases, the wave is separated
into regions with the right and left circular polarizations. 
}
\label{separation} 
\end{figure}

\section{CONCLUSIONS}

A model has been proposed and verified as a useful working tool to study strictly
monochromatic nonlinear waves. This method will allow a significant advance in a
number of interesting problems. It would be reasonable to find explicitly 
conservative dynamic systems close to our model in order to better understand 
problems of stability of small-scale structures.

The problem of the spatiotemporal development of the obtained stationary patterns
under small nonmonochromatic perturbations requires a separate study. In particular,
it is still unclear to which distance the self-focused thin beam can propagate under 
the action of chromatic dispersion. In this context, a special attention should also 
be paid to quasi-cylindrical structures with domain walls because they tend to be 
separated into three-dimensional ``drops'' in the case of the anomalous dispersion
of the group velocity [37]. However, since characteristic longitudinal scales (about
hundreds and thousands of wavelengths) in our numerical experiments appeared to be
not too large compared to the available coherence length of the incident laser beam,
it can be hopped that the stationary pattern at a sufficiently large coherence length
does not become worse at such distances.

\vspace{2mm}

{\bf FUNDING}. This work was supported by the Ministry of Science and Higher 
Education of the Russian Federation (state assignment no. FFWR-2024-0013).

\vspace{2mm}

{\bf CONFLICT OF INTEREST}.
The author of this work declares that he has no conflicts of interest.

\end{document}